\documentclass[11pt]{amsart}

\usepackage[utf8]{inputenc}
\usepackage[T1]{fontenc}
\usepackage{lmodern}

\usepackage{amssymb}
\usepackage{mathtools}
\usepackage{amscd}
\usepackage[mathscr]{eucal}
\usepackage{enumitem}
\usepackage[hmargin=1in,vmargin=1in]{geometry}

\usepackage{graphicx}
\usepackage{xcolor}

\usepackage[roman]{sublabel}

\usepackage[labelfont=bf,font=small]{caption}
\usepackage{subcaption}

\usepackage[noadjust]{cite}
\usepackage[foot]{amsaddr}

\usepackage{placeins}
\usepackage{bm}
\usepackage{siunitx}

\definecolor{cite_purple}{RGB}{128,9,158}
\definecolor{cite_blue}{RGB}{2,95,176}
\definecolor{link_red}{rgb}{0.7,0,0}
\usepackage[colorlinks=true,
            citecolor=cite_purple,
            linkcolor=link_red,
            urlcolor=cite_blue]{hyperref}

\newlength{\arrayrulewidthOriginal}

\theoremstyle{plain}

\theoremstyle{definition}

\begin{document}

\title{Navigating the Delicate Geometry of Beehive Mite Infestation with Optimal Control}

\author{Julia Saff$^{1}$}
\address{$^{1}$Anne Spencer Daves
College of Education, Health, and Human Sciences, Florida State University, Tallahassee, Florida, 32306, USA}

\author{Bhargav R. Karamched$^{*,2,3,4}$}
\address{$^{2}$ Department of Mathematics, Florida State University, Tallahassee, Florida 32306, USA}
\address{$^{3}$Institute of Molecular Biophysics, Florida State University, Tallahassee, Florida 32306, USA}
\address{$^4$ Program in Neuroscience, Florida State University, Tallahassee, Florida 32306, USA}
\thanks{$^{*}$\href{mailto:bkaramched@fsu.edu}{bkaramched@fsu.edu} (Corresponding author)}



\date{\today}

\begin{abstract}
The parasitic mite \textit{Varroa destructor} poses a severe existential threat to global honey bee (\textit{Apis mellifera}) populations. In this paper, we present a dynamical systems model of hive-mite interactions incorporating a eusocial Allee effect to evaluate the efficacy of chemical interventions. We partition treatments into ``soft'' miticides (targeting phoretic mites) and ``harsh'' miticides (penetrating the capped brood to target reproductive mites). Our bifurcation analysis reveals a fundamental trade-off: while harsh treatments effectively eradicate the protected mite reservoir by shifting the transcritical bifurcation boundary, they impose sublethal toxicity on the bees. We analytically show that exceeding a critical dosage threshold culminates in a catastrophic saddle-node bifurcation that guarantees colony collapse. To navigate this toxicity limit, we formulate an optimal control problem using Pontryagin's Maximum Principle. Numerical solutions reveal that a dynamic ``shock and maintain'' cocktail strategy optimally balances reservoir clearance with hive viability. Expanding the model to include treatment-resistant strains demonstrates that single-chemical reliance forces the hive onto a highly toxic, nearly unsustainable chemical treadmill. Finally, our mathematical framework yields a striking ecological prediction: because the eradication boundary scales intimately with the colony's carrying capacity, unchecked mite pressure will likely exert evolutionary forces that select for smaller, naturally resistant hives over the massive colonies favored by commercial agriculture. 
\end{abstract}

\maketitle

\raggedbottom
\thispagestyle{empty}

\section{Introduction}

A healthy bee hive reflects how bees are masters of organization, communication, and construction engineering~\cite{sammataro2000parasitic,johnson2003organization}. In the insect world, they are a true model of efficiency which has allowed them to successfully propagate worldwide and survive in nearly all environments~\cite{seeley2011honeybee, hrncir2019stingless}. Their existence is essential to pollination and the maintenance of the human food chain~\cite{khalifa2021overview}. Unfortunately, the introduction of a small mite named the \textit{Varroa destructor} in the late 20th century to the Western honeybee, \textit{Apis mellifera}, has caused widespread destruction and colony collapse throughout North America and Europe~\cite{traynor2020varroa}. This is of great concern not only to honey production, but also to successful pollination in the plant world. These mites are parasitic and attach directly onto the bees, using them as a food source in addition to being a vector of a deadly virus which destroys their developing offspring~\cite{jeyapriya2025parasitic}. 

As a result of this pathogen, there has been a race to find the most effective treatment to manage and minimize the existence of \textit{Varroa}, though their total obliteration is unlikely. The choice of protocols depends upon the environment, outdoor temperature, season, development of treatment resistance, and the beekeeper’s preference for a more organic versus synthetic chemical approach~\cite{underwood2003effects,jack2024evaluating,rinkevich2020detection,thurston2025efficacy, zufriategui2024detrimental}. Amitraz, a chemical miticide, can be helpful, but there is evidence of recent mite resistance to this drug~\cite{bertola2025sensitivity}. Additionally, in too strong a dose, it can harm bees~\cite{kayode2014effect}. Its use is restricted to the season when the bees are not making honey for human consumption~\cite{richards2021honey, chaimanee2022determination}. Other more organic treatments, such as formic or oxalic acid, can be used; these have no deleterious effects if they get into consumed honey but need to be applied during an ideal time and temperature~\cite{bogdanov2002determination}. 

Mechanical techniques also exist, such as utilizing a screened bottom board that allows mites to fall out from the hive’s floor~\cite{liu2020meta}, or removing the male drone brood, which tends to harbor a larger reservoir of mites~\cite{calderone2005evaluation}. Season, temperature, past success in local hives, colony size, and the temperament of the beekeeper all play a role in treatment selection~\cite{jack2023seasonal,underwood2003effects,van2021can,giacobino2016key}. Some bee experts claim that keeping the colony large and well-nourished with supporting vitamins is sufficient for the colony to fight off the mites~\cite{alaux2011nutrigenomics, glavinic2017dietary}. Still others hope mite-resistant bee strains via genetic mutations will eventually evolve with improved colony resistance~\cite{buchler2010auslese,rinderer2010breeding}. Considering all of the above, an uncontroversial statement is that no two beekeepers believe in the exact same approach to mitigating mite infestation.

Mathematical modeling is a powerful tool that can provide crucial insights in complex biological scenarios exactly like this. Namely, it can capture the essential physical processes underlying bee population health and make predictions on how to impute treatments to maintain a resilient hive. Modeling efforts in the context of honeybees have predominantly focused on foraging strategies and the mechanistic underpinnings of the remarkably sophisticated information cascade that exists in colonies~\cite{bidari2021hive, khoury2013modelling, camazine1990mathematical,bagheri2019mathematical}. In the context of mite infestation, modeling has focused on the impact upon bee population dynamics~\cite{degrandi2004mathematical, martin2001role,ratti2017mathematical} and has provided insight on when the application of treatments can be most influential without harming hive health~\cite{torres2015modeling, martin1998population, schodl2022simulation}. 

In this paper, we introduce a model of honeybee and mite population dynamics to investigate how a cocktail of harsh treatments (e.g., Amitraz or formic acid) and softer treatments (e.g., oxalic acid) can be optimally implemented. We parametrize our model using established literature on \textit{Varroa} mites and invoke optimal control theory~\cite{lenhart2007optimal} to predict an adaptive treatment protocol.

Our goal is not the complete obliteration of the mite population. Although desirable, attempting total eradication with intensive treatment protocols severely risks killing the honeybees themselves via chemical toxicity. Instead, we focus on optimal mitigation, framing our objective functional to manage rather than annihilate the parasite burden. We borrow this philosophy from adaptive therapy protocols in cancer treatment~\cite{west2020towards,wodarz2021adaptive,wodarz2008use,gallagher2026mathematical} and the management of antibiotic-resistant bacterial infections~\cite{dere2025optimal}, which have shown immense promise as intervention structures for maintaining long-term system health.

To realistically capture this dynamic, we explicitly partition the mite populations into phoretic and reproductive phases. Phoretic mites are adult parasites that attach directly to adult bees, using them for sustenance and transport throughout the colony. In contrast, reproductive mites are those that have invaded and become sealed inside capped brood cells, where they feed on developing bee pupae and actively multiply to create a protected reservoir. Because of this physical wax barrier, reproductive mites are unaffected by soft treatments. We further incorporate the emergence of miticide-resistant strains---an increasingly challenging problem that puts colonies at severe risk~\cite{rodriguez2005resistance, rinkevich2020detection}. By tracking mite reproduction within capped brood cells, our model provides mechanistic insights into how the mite reservoir drives colony mortality. 

Furthermore, recent ecological modeling establishes that obligate social insects like honey bees exhibit a strong Allee effect~\cite{stephens1999what}: a critical minimum population threshold is required for adequate brood care, thermoregulation, social cohesion, and hive survival~\cite{eberl2010importance, khoury2011quantitative}. If environmental stressors or heavy parasite loads push the adult bee population below this critical size, the colony loses resilience and inevitably collapses~\cite{perry2015rapid, betti2014effects}. While previous studies have utilized mathematical epidemiology to predict disease transmission and eventual colony failure~\cite{kang2016disease, ratti2015mathematical}, there remains a critical need to integrate optimal control theory directly with these Allee models. Our model explicitly incorporates the Allee effect in honeybee population dynamics. By framing the administration of miticides as a dynamic control problem, we derive treatment trajectories that actively suppress both susceptible and resistant mite reservoirs while dynamically preventing chemical toxicity from pushing the fragile colony below its collapse threshold.



\section{Model and Analysis}
Let $B(t)$, $M_P(t)$, and $M_R(t)$ represent the population of bees, phoretic mites, and reproductive mites, respectively, as functions of time. These state variables evolve according to the dynamical equations
\begin{align}
\begin{split}
    \frac{dB}{dt} &= \alpha B(B-A)(K-B) - \beta BM_P\\
    \frac{dM_P}{dt} &= \nu\sigma M_R - \gamma B M_P - \delta M_P\\
    \frac{dM_R}{dt} &= -\sigma M_R + \gamma BM_P,
    \end{split}
    \label{eq:model1}
\end{align}
where $\alpha$ is the growth rate of bees, $\beta$ represents the killing of bees by phoretic mites, $\sigma$ represents a brood emergence rate, $\gamma$ represents a brood invasion rate, and $\delta$ represents the natural death of phoretic mites. We assume logistic growth with the Allee effect to describe bee growth because typically in natural settings healthy hives require a minimum number of bees to be present. This critical threshold population is denoted by $A$. The carrying capacity of the hive is denoted $K$. 

For the baseline parameter values used in our numerical simulations, see Table~\ref{tab:parameters}. Values governing the natural population dynamics of the bees and mites are derived from established apicultural literature. Because previous literature often utilizes higher-dimensional compartment or step-simulation models~\cite{khoury2011quantitative,martin1998population,ratti2015mathematical}, the parameter values in Table \ref{tab:parameters} defining our low-dimensional macroscopic ODEs were phenomenologically calibrated to match the timescale and population thresholds established in these foundational works.

\begin{table}[htbp]
    \centering
    \caption{Baseline Parameter Values}
    \label{tab:parameters}
    \renewcommand{\arraystretch}{1.3} 
    \begin{tabular}{clcl}
        \hline
        \textbf{Parameter} & \textbf{Interpretation} & \textbf{Value} & \textbf{Reference} \\
        \hline
        $K$ & Hive carrying capacity & $50,000$ & Estimated from \cite{khoury2011quantitative} \\
        $A$ & Allee effect critical threshold & $5,000$ & Estimated from \cite{eberl2010importance, khoury2011quantitative} \\
        $\alpha$ & Bee population growth scalar & $1.5 \times 10^{-9}$ & Calibrated from \cite{khoury2011quantitative} \\
        $\beta$ & Parasite-induced bee mortality rate & $5 \times 10^{-5}$ & Calibrated from \cite{ratti2015mathematical} \\
        $\nu$ & Mite reproduction factor & $1.5$ & \cite{martin1998population} \\
        $\sigma$ & Brood emergence rate & $0.077$ & \cite{martin1998population} \\
        $\gamma$ & Brood invasion rate & $1 \times 10^{-5}$ & Calibrated from \cite{martin1998population} \\
        $\delta$ & Phoretic mite natural death rate & $0.01$ & \cite{martin1998population} \\
        \hline
    \end{tabular}
\end{table}

\subsection{Equilibria and Linear Stability Analysis}

To understand the baseline dynamics of the hive, we determine the equilibria of the system and evaluate their local stability. We set the time derivatives in Eq.~\eqref{eq:model1} to zero, yielding the following algebraic equations for the equilibria of the system:

\begin{align}
    &\alpha B(B-A)(K-B) - \beta B M_P=0, \label{eq:B_steady} \\
    &\nu\sigma M_R - \gamma B M_P - \delta M_P=0, \label{eq:Mp_steady} \\
    &-\sigma M_R + \gamma B M_P = 0. \label{eq:Mr_steady}
\end{align}

From \eqref{eq:Mr_steady}, we find the relationship between reproductive and phoretic mites at equilibrium:
\begin{equation}
    M_R = \frac{\gamma B}{\sigma} M_P. \label{eq:Mr_sub}
\end{equation}
Substituting \eqref{eq:Mr_sub} into \eqref{eq:Mp_steady} yields:
\begin{equation}
    0 = M_P \left[ \gamma B (\nu - 1) - \delta \right].
\end{equation}
This gives rise to two biological classifications of equilibria: disease-free equilibrium (where $M_P = 0$) and an endemic (coexistence) equilibrium (where $M_P > 0$).

If $M_P = 0$, then $M_R = 0$. Substituting $M_P = 0$ into \eqref{eq:B_steady} leaves $\alpha B(B-A)(K-B) = 0$, giving three distinct disease-free states:
\begin{enumerate}
    \item \textbf{Colony Collapse:} $E_0 = (0, 0, 0)$
    \item \textbf{Allee Threshold:} $E_A = (A, 0, 0)$
    \item \textbf{Healthy Hive:} $E_K = (K, 0, 0)$\\
\end{enumerate}

\noindent\textbf{{Endemic Equilibrium}.} Assuming an active mite infestation ($M_P \neq 0$), we divide by $M_P$ to find the critical bee population required to sustain the mites:
\begin{equation}
    B^* = \frac{\delta}{\gamma(\nu - 1)}.
\end{equation}
Substituting $B^*$ into \eqref{eq:B_steady} and assuming $B^* \neq 0$, we find the steady-state phoretic mite population:
\begin{equation}
    M_P^* = \frac{\alpha}{\beta} (B^* - A)(K - B^*).
\end{equation}
The reproductive mite population follows as $M_R^* = \frac{\gamma B^*}{\sigma} M_P^*$. This endemic equilibrium, denoted $E_{coex} = (B^*, M_P^*, M_R^*)$, is biologically relevant only if $M_P^* > 0$, which necessitates $A < B^* < K$.\\

\noindent\textbf{Linear Stability Analysis.} To determine local asymptotic stability, we compute the Jacobian matrix $J$ of the system:
\begin{equation}
    J = \begin{bmatrix}
        \alpha(-3B^2 + 2(A+K)B - AK) - \beta M_P & -\beta B & 0 \\
        -\gamma M_P & -\gamma B - \delta & \nu\sigma \\
        \gamma M_P & \gamma B & -\sigma
    \end{bmatrix}.
\end{equation}

Evaluating $J$ at $E_0 = (0,0,0)$ yields a diagonal matrix with eigenvalues $\lambda_1 = -\alpha AK$, $\lambda_2 = -\delta$, and $\lambda_3 = -\sigma$. Because all parameters are strictly positive, all eigenvalues are strictly negative, rendering $E_0$ a stable node. Thus, any small perturbation away from the origin will exponentially degrade away and return to extinction. This is characteristic of the Allee effect: a collapsed hive cannot spontaneously recover.

Evaluating $J$ at $E_A = (A,0,0)$ yields $\lambda_1 = \alpha A(K-A)$. Since $K > A$, this eigenvalue is strictly positive, rendering the Allee threshold $E_A$ unstable (a saddle point). Thus, any small perturbation away from $E_A$ will exponentially diverge away from $E_A$, except along the eigendirections specified by the negative eigenvalues. We do not consider perturbations in such specialized directions as generic. 

\begin{figure}[htbp]
    \centering
    \includegraphics[width=0.9\linewidth]{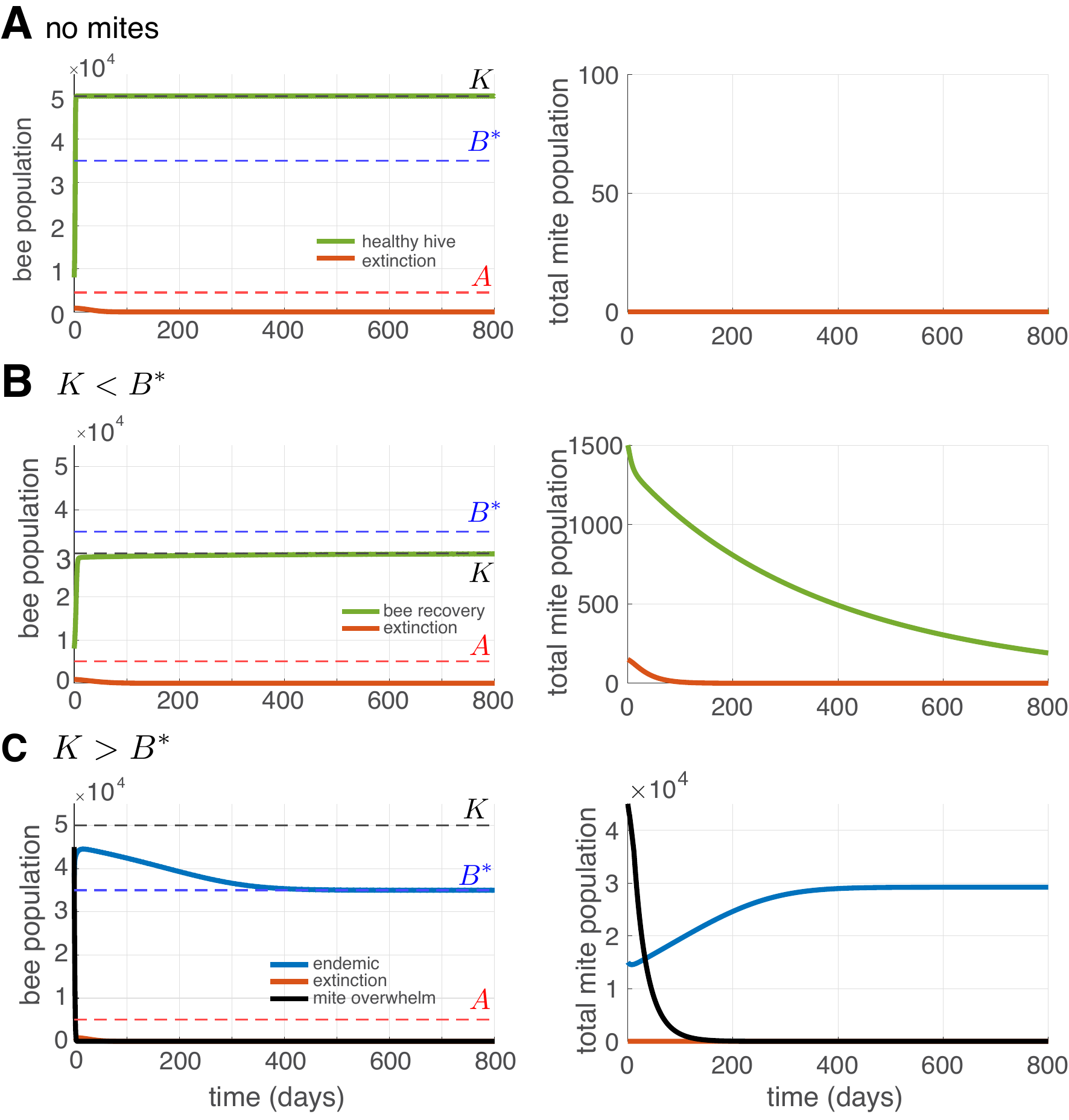}
    \caption{Time series for bees and mites for (A) the no mite case, (B) the case where $K < B^*$, showing the hive is not large enough to sustain mite population, and (C) the case where $K > B^*$, showing endemic state. In this case, if the mite population gets too large, they overwhelm the dynamics and cause bee extinction (black). Parameters are as in Table~\ref{tab:parameters}.}
    \label{fig:fig1}
\end{figure}

\begin{figure}[htbp]
    \centering
    \includegraphics[width=0.6\linewidth]{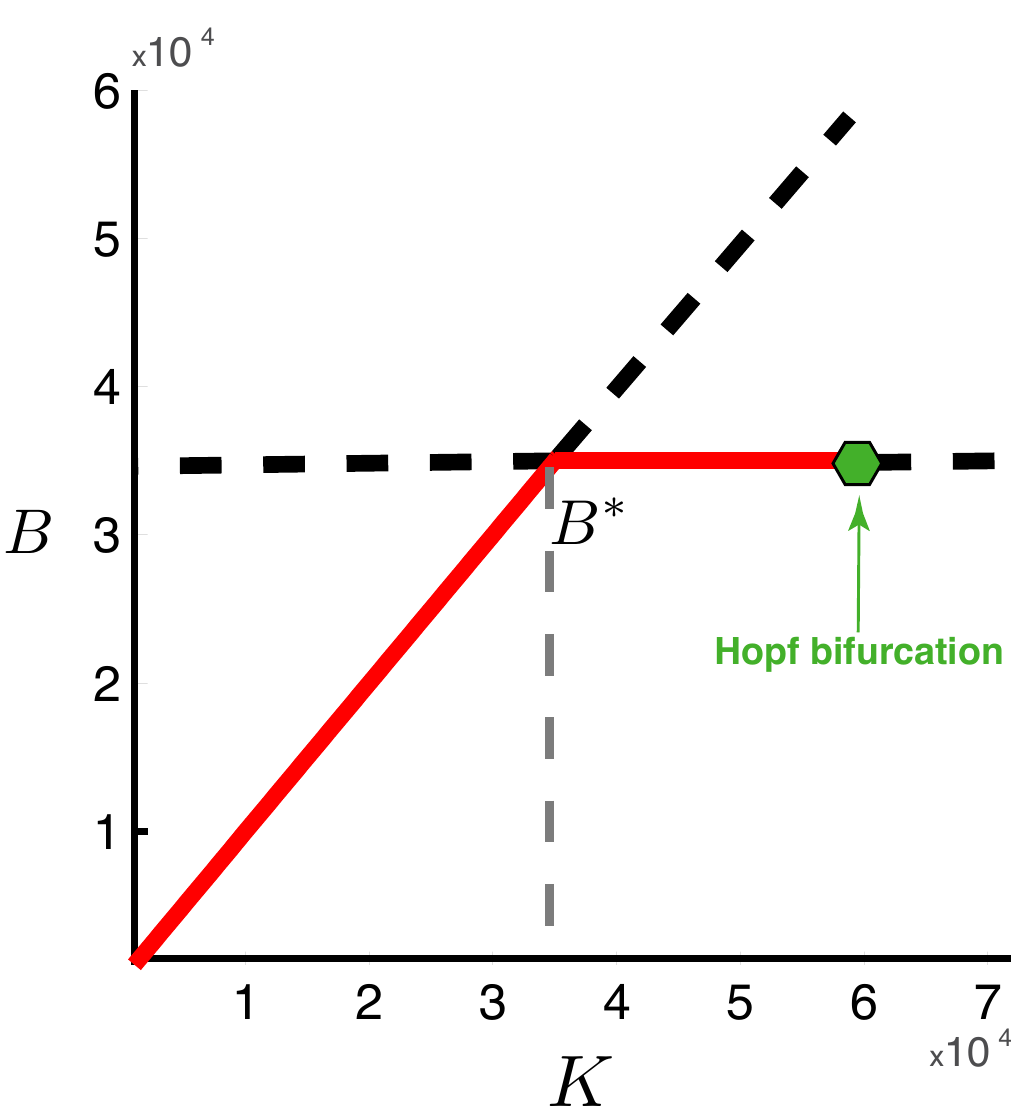}
    \caption{Bifurcation diagram for Eq~\eqref{eq:model1}. A transcritical bifurcation occurs at $K = B^*$ providing a criteria by which the endemic equilibrium is stabilized. Moreover, there is a subcritical Hopf bifurcation yielding a homoclinic connection, resulting in a collapse to extinction.}
    \label{fig:fig2}
\end{figure}

Evaluating $J$ at the healthy state $E_K = (K,0,0)$ yields $\lambda_1 = -\alpha K(K-A) < 0$. The stability is thus determined by the remaining $2 \times 2$ submatrix governing the mite dynamics:
\begin{equation}
    J_{mites}(E_K) = \begin{bmatrix}
        -\gamma K - \delta & \nu\sigma \\
        \gamma K & -\sigma
    \end{bmatrix}.
\end{equation}
The trace is strictly negative: $\text{Tr} = -(\gamma K + \delta + \sigma) < 0$. Stability requires a positive determinant:
\begin{align}
    &\sigma\delta + \sigma\gamma K(1 - \nu) > 0\nonumber\\
    &\implies K < \frac{\delta}{\gamma(\nu - 1)} = B^*.
\end{align}
Thus, $E_K$ is locally asymptotically stable if $K < B^*$, meaning the hive cannot support the mites. If $K > B^*$, $E_K$ loses stability, and a small introduction of mites will inevitably lead to an infestation. We demonstrate this numerically in Figure~\ref{fig:fig1}, where the endemic equilibrium indeed stabilizes as $K$ grows past $B^*$. Dynamically, this occurs via a transcritical bifurcation (see Figure~\ref{fig:fig2}). 
 
 Interestingly, as $K$ is increased significantly beyond the transcritical establishment threshold $B^*$, numerical continuation reveals that the endemic equilibrium loses stability via a Hopf bifurcation (at $K \approx 60,000$ for baseline parameters). Thus, in artificially massive hives, the extreme abundance of host bees allows the mite population to aggressively over-reproduce, driving population oscillations. However, continuation of the periodic branch reveals no stable limit cycles; rather, the branch undergoes a rapid sequence of destabilizing bifurcations (including limit points and period doublings). Geometrically, the amplitude of the population cycles grows so large that the downward swing of the bee population breaches the Allee threshold $A$. This collision with the unstable saddle creates a homoclinic bifurcation, instantly destroying the periodic orbit and guaranteeing hive collapse. This is a rich area that we hope to investigate in future work, particularly pertaining to the biological consequences.

 The above analysis yields a critical biological insight: mite establishment fundamentally requires a sufficiently large host population to persist. If a honeybee colony maintains a naturally small but sustainable population size ($A < K < B^*$), a mite infestation cannot take hold.

However, relying on small colonies is economically unviable. Commercial apiculture depends on massive honey surpluses, which in turn require an artificially large foraging workforce. Beekeepers actively maintain hive carrying capacities at extreme levels ($K \approx 50,000$), placing the typical managed colony well beyond the $B^*$ threshold dictated by mite ecological constraints~\cite{lazutin2020keeping,seeley2011honeybee}. Consequently, the economic imperative to maximize honey production inherently traps managed hives in a state of continuous vulnerability to mite infestation, necessitating the use of chemical controls. Furthermore, the existence of the chaotic Hopf bifurcation underscores the fragility of artificially large beehives. We next investigate the impact of individual treatments upon bee and mite dynamics.

\section{Analysis of Isolated Treatment Strategies}
\label{sec:iso_treat}
To motivate the necessity of a dynamic "cocktail" approach, we first evaluate the efficacy of two standard, isolated treatment strategies: a soft phoretic treatment (e.g., oxalic acid vapor) and a harsh reproductive treatment (e.g., Amitraz). We incorporate these treatments into the baseline model as constant control parameters, $u_1$ and $u_2$, respectively. 

\subsection{Case 1: Soft Treatment}
The soft treatment $u_1$ selectively increases the mortality of phoretic mites ($M_P$) while exhibiting negligible toxicity to the bees. The modified system becomes:
\begin{align}
    \frac{dB}{dt} &= \alpha B(B-A)(K-B) - \beta B M_P, \label{eq:soft_B} \\
    \frac{dM_P}{dt} &= \nu\sigma M_R - \gamma B M_P - (\delta + u_1) M_P, \label{eq:soft_Mp} \\
    \frac{dM_R}{dt} &= -\sigma M_R + \gamma B M_P. \label{eq:soft_Mr}
\end{align}
The presence of $u_1$ does not alter the location of the disease-free states ($E_0, E_A, E_K$). However, it fundamentally shifts the endemic coexistence equilibrium. Following the same algebraic reduction as the untreated model, the new critical bee population required to sustain the infestation becomes:
\begin{equation}
    B^*_{soft} = \frac{\delta + u_1}{\gamma(\nu - 1)}.
\end{equation}
Because $u_1 > 0$, the required bee population $B^*_{soft}$ is strictly greater than the untreated baseline $B^*$. Biologically, the treatment makes it harder for mites to find enough hosts to overcome their new, elevated death rate.

\begin{figure}[htbp]
    \centering
    \includegraphics[width=0.6\linewidth]{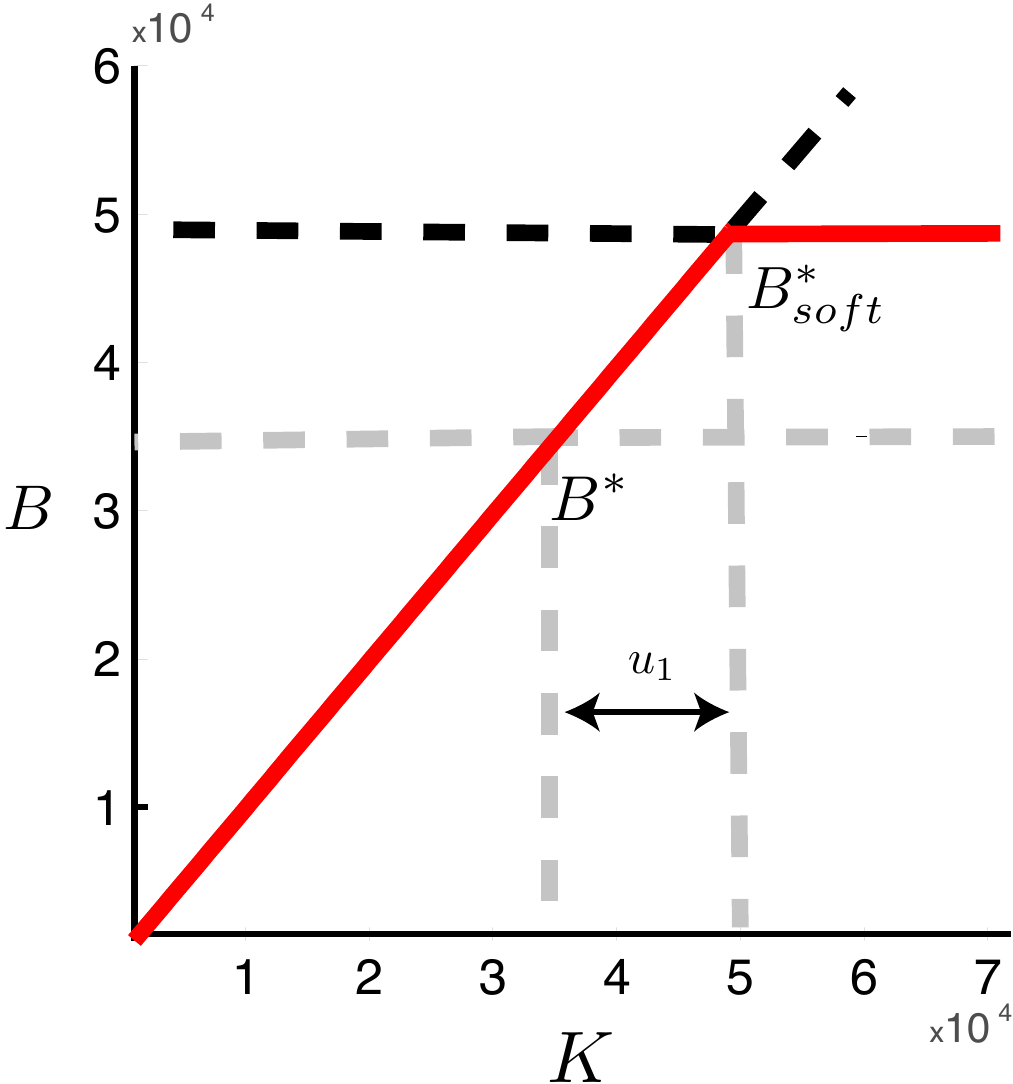}
    \caption{Bifurcation diagram for Eq~\eqref{eq:model1} with the soft treatment implemented. The result is that the transcritical bifurcation point moves to the right and allows for larger hives to exist without the mite infestation.}
    \label{fig:fig3}
\end{figure}

In the regime that $B^*$ is stable, as $u_1$ is introduced and increased, $B^*_{soft}$ moves closer to the carrying capacity $K$. If the treatment intensity is sufficiently high such that $u_1 > \gamma K(\nu - 1) - \delta$, we force $B^*_{soft} > K$. At this exact critical threshold, the system undergoes the same transcritical bifurcation as discussed previously (see Figure~\ref{fig:fig3}). The endemic state exits the biologically feasible region, and the healthy state $E_K$ becomes a stable node, theoretically eradicating the mites. Another interpretation of this is that the soft treatment allows for larger hives to exist without the threat of a mite infestation.

However, while mathematically capable of eradication, this strategy is practically suboptimal due to the reservoir effect. Notice that the control $u_1$ is absent from the reproductive mite equation \eqref{eq:soft_Mr}. The capped brood acts as a protected reservoir, constantly releasing new mites ($\nu\sigma M_R$) into the phoretic compartment. To maintain $B^*_{soft} > K$, the beekeeper must apply $u_1$ continuously and indefinitely. If the continuous treatment is ceased or constrained by labor costs, the system immediately drops back below the transcritical bifurcation point, trapping the colony in a suppressed, yet chronic, endemic state. 

\subsection{Case 2: Harsh Treatment}
\begin{figure}[htbp]
    \centering
    \includegraphics[width=\linewidth]{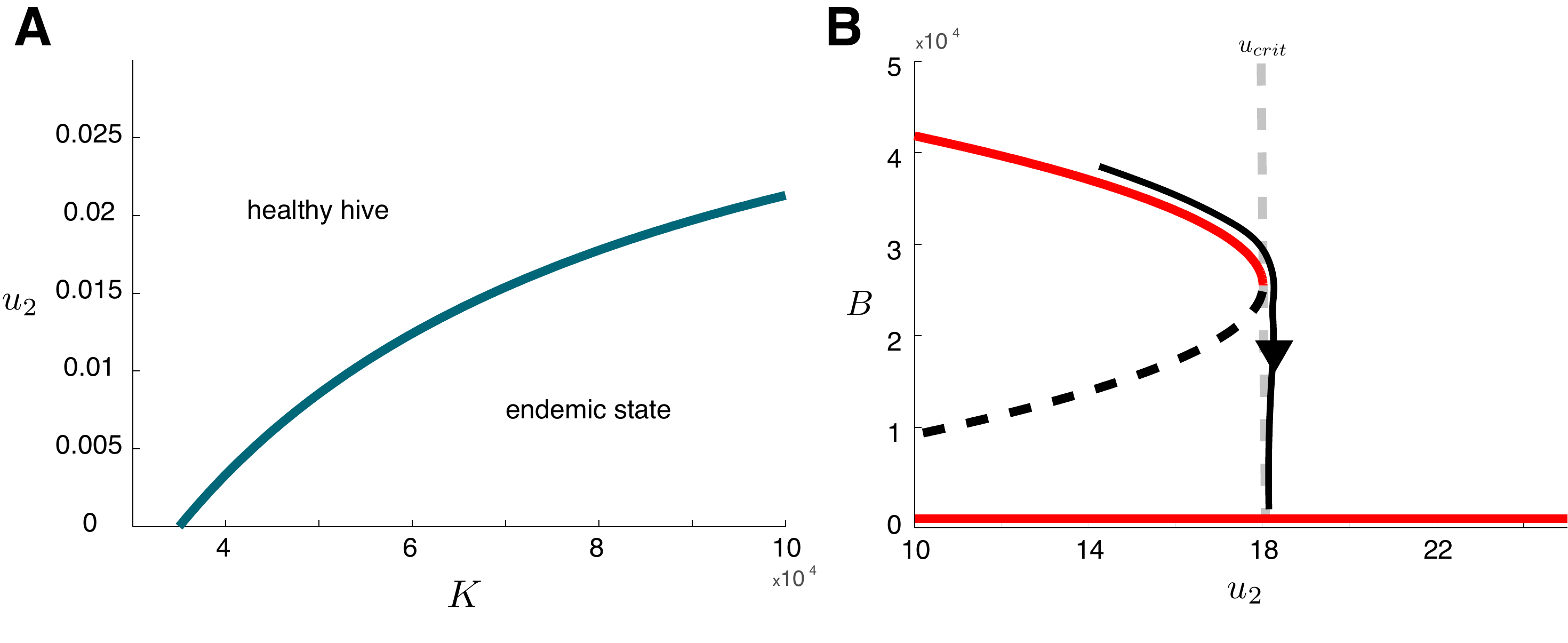}
    \caption{Dynamical profile for the harsh treatment case. (A) Locus of transcritical bifurcation points. The harsh treatment functionally increases the hive size that can be maintained in a healthy fashion, similar to the effect of $u_1$, but the impact is nonlinear. (B) Intensely using the harsh treatment pushes the system past an irreversible saddle-node bifurcation point.}
    \label{fig:fig4}
\end{figure}

To target the hidden reservoir, beekeepers employ harsh treatments that penetrate capped cells to kill reproductive mites ($M_R$). However, these chemicals incur a severe physiological cost to the bees, modeled as an additional mortality rate $\eta u_2 B$. The system for the harsh treatment only case is:
\begin{align}
    \frac{dB}{dt} &= \alpha B(B-A)(K-B) - \beta B M_P - \eta u_2 B, \label{eq:harsh_B} \\
    \frac{dM_P}{dt} &= \nu\sigma M_R - \gamma B M_P - \delta M_P, \label{eq:harsh_Mp} \\
    \frac{dM_R}{dt} &= -\sigma M_R + \gamma B M_P - u_2 M_R. \label{eq:harsh_Mr}
\end{align}
From \eqref{eq:harsh_Mr}, the equilibrium ratio of reproductive to phoretic mites is suppressed compared to the untreated case: $M_R = \frac{\gamma B}{\sigma + u_2} M_P$. Substituting this into \eqref{eq:harsh_Mp} yields the new critical bee density for the endemic state:
\begin{equation}
    B^*_{harsh} = \frac{\delta}{\gamma \left( \frac{\nu\sigma}{\sigma + u_2} - 1 \right)}.
\end{equation}
Functionally, the harsh treatment works the same way as the soft treatment, namely by increasing the hive size that can be maintained without a mite infestation. This manifests through a shifted transcritical bifurcation point. In Figure~\ref{fig:fig4}A, we show how the transcritical bifurcation point shifts in $(u_2, K)$ parameter space. Importantly, introducing the harsh treatment allows for the maintenance of a healthy hive.
Based on this finding, one may ask: why not simply drench the hive with the harsh treatment? While $u_2$ effectively restricts mite growth, the toxicity term in the bee equation radically alters the disease-free states. Factoring out $B$ in \eqref{eq:harsh_B} when $M_P = 0$, the new healthy state and Allee threshold are the roots of the quadratic:
\begin{equation}
    \alpha(B-A)(K-B) - \eta u_2 = 0.
    \label{eq:down_parabola}
\end{equation}
This is a downward-facing parabola shifted downward by the toxicity $\eta u_2$. As treatment intensity $u_2$ increases, the roots of the quadratic become closer together, meaning the effective carrying capacity ($K_{eff}$) and effective Allee threshold ($A_{eff}$) converge. 

At a critical treatment intensity $u_{2,crit}$, the discriminant of this quadratic becomes zero:
\begin{equation}
    u_{2,crit} = \alpha \frac{(K-A)^2}{4\eta}
    \label{eq:u2crit}
\end{equation}
Here, the system undergoes a catastrophic saddle-node bifurcation. The disease-free state ($E_K$) and the Allee threshold ($E_A$) collide and mutually annihilate (see Figure~\ref{fig:fig4}B). 

The mathematical implication of this saddle-node bifurcation is dire. If $u_2$ slightly exceeds $u_{2,crit}$, or if the hive's population is already marginalized by prior mite damage, the treatment pushes the colony across the shifting Allee threshold. The only remaining equilibrium in the system is $E_0 = (0,0,0)$. Even worse, the saddle-node bifurcation manifests a one-way switch, meaning simply returning $u_2$ values to those where a healthy hive was observed does nothing to help. The attempt to cure the hive guarantees its collapse.

\subsection{Motivation for Optimal Control}
The dynamical analysis of the isolated treatments reveals a fundamental trade-off. The soft treatment ($u_1$) safely manages the symptoms but fails to clear the reservoir, leading to an intractable endemic trap. The harsh treatment ($u_2$) attacks the root of the infestation but risks saddle-node annihilation via chemical toxicity. Moreover, combining the two treatments in a static implementation does not resolve the sensitivity of the hive health to treatments, as the catastrophic saddle-node bifurcation continues to exist.

To simultaneously eradicate the parasite reservoir and preserve the social integrity of the hive, we must abandon static treatment paradigms. In the following section, we formulate an optimal control problem to derive a dynamic, time-dependent "cocktail" trajectory $(u_1^*(t), u_2^*(t))$ that navigates the narrow manifold between the endemic trap and the Allee collapse threshold.

\section{Optimal Control of the Treatment Cocktail}

Given the established suboptimality and collapse risks associated with static treatment strategies, we formulate an optimal control problem to dynamically orchestrate the soft ($u_1(t)$) and harsh ($u_2(t)$) treatments. The goal is to maximize the bee population while minimizing both the parasitic load and the physiological/economic costs of the treatments throughout the time interval of interest $[0, T]$.

We define the objective functional $J(u_1, u_2)$ to be maximized:
\begin{equation}
    J(u_1, u_2) = w_1 B(T) - \int_0^T \left( w_2 M_P(t) + w_3 M_R(t) + C_1 u_1^2(t) + C_2 u_2^2(t) \right) dt. \label{eq:objective}
\end{equation}
The terminal payoff $w_1 B(T)$ heavily weights the survival of the colony at the end of the time interval. Inside the integral, $w_2$ and $w_3$ represent the ongoing damage inflicted by the phoretic and reproductive mite reservoirs, respectively. The quadratic terms $C_1 u_1^2$ and $C_2 u_2^2$ serve a dual purpose: they represent the nonlinear economic and labor costs of applying treatments, and mathematically, they discourage biologically unrealistic bang-bang impulses of infinite magnitude~\cite{lenhart2007optimal}.

We seek an optimal control pair $(u_1^*, u_2^*)$ such that:
\begin{equation}
    J(u_1^*, u_2^*) = \max_{\substack{0 \le u_1 \le u_{1,max} \\ 0 \le u_2 \le u_{2,max}}} J(u_1, u_2)
\end{equation}
subject to the state dynamics derived in the combined model:
\begin{align}
    \frac{dB}{dt} &= \alpha B(B-A)(K-B) - \beta B M_P - \eta u_2 B, \label{eq:state_B} \\
    \frac{dM_P}{dt} &= \nu\sigma M_R - \gamma B M_P - (\delta + u_1) M_P, \label{eq:state_Mp} \\
    \frac{dM_R}{dt} &= -(\sigma + u_2) M_R + \gamma B M_P, \label{eq:state_Mr}
\end{align}
with initial conditions $B(0) = B_0$, $M_P(0) = M_{P0}$, and $M_R(0) = M_{R0}$.

We employ Pontryagin's Maximum Principle to determine $(u_1^*, u_2^*)$~\cite{lenhart2007optimal}. To do so, we convert the constrained optimization problem into the Hamiltonian $H$. We introduce the time-dependent adjoint variables (or co-states) $\lambda_1(t), \lambda_2(t)$, and $\lambda_3(t)$, which represent the dynamic shadow prices (marginal value) of the bees, phoretic mites, and reproductive mites, respectively.

The Hamiltonian is defined as the integrand of the objective functional plus the inner product of the adjoint vector and the state dynamics:
\begin{align}
    H = &-(w_2 M_P + w_3 M_R + C_1 u_1^2 + C_2 u_2^2) \nonumber \\
    &+ \lambda_1 \Big[ \alpha B(B-A)(K-B) - \beta B M_P - \eta u_2 B \Big] \nonumber \\
    &+ \lambda_2 \Big[ \nu\sigma M_R - \gamma B M_P - (\delta + u_1) M_P \Big] \nonumber \\
    &+ \lambda_3 \Big[ -(\sigma + u_2) M_R + \gamma B M_P \Big]. \label{eq:hamiltonian}
\end{align}

The dynamic evolution of the adjoint variables is governed by the differential equations $\dot{\lambda}_i = -\frac{\partial H}{\partial x_i}$, which yields:
\begin{align}
    \frac{d\lambda_1}{dt} &= -\lambda_1 \Big[ \alpha(-3B^2 + 2(A+K)B - AK) - \beta M_P - \eta u_2 \Big] + \lambda_2 (\gamma M_P) - \lambda_3 (\gamma M_P), \label{eq:adjoint_1} \\
    \frac{d\lambda_2}{dt} &= w_2 + \lambda_1 (\beta B) + \lambda_2 (\gamma B + \delta + u_1) - \lambda_3 (\gamma B), \label{eq:adjoint_2} \\
    \frac{d\lambda_3}{dt} &= w_3 - \lambda_2 (\nu\sigma) + \lambda_3 (\sigma + u_2). \label{eq:adjoint_3}
\end{align}
Because the terminal payoff in $J$ only explicitly rewards the bee population $B(T)$, the transversality conditions for the adjoint system at the final time $t = T$ are given by:
\begin{equation}
    \lambda_1(T) = w_1, \quad \lambda_2(T) = 0, \quad \lambda_3(T) = 0. \label{eq:transversality}
\end{equation}

\subsection{Characterization of the Optimal Controls}
The optimal controls $u_1^*(t)$ and $u_2^*(t)$ must maximize the Hamiltonian $H$ at almost every time $t \in [0, T]$. We differentiate $H$ with respect to each control variable and set the derivatives to zero:
\begin{align}
    \frac{\partial H}{\partial u_1} &= -2 C_1 u_1 - \lambda_2 M_P = 0, \\
    \frac{\partial H}{\partial u_2} &= -2 C_2 u_2 - \lambda_1 \eta B - \lambda_3 M_R = 0.
\end{align}
Solving for the controls, and accounting for the defined biological bounds $[0, u_{max}]$, we obtain the piecewise characterization for the optimal "cocktail" strategy:
\begin{align}
    u_1^*(t) &= \min \left\{ \max \left\{ 0, \frac{-\lambda_2 M_P}{2 C_1} \right\}, u_{1,max} \right\}, \label{eq:opt_u1} \\
    u_2^*(t) &= \min \left\{ \max \left\{ 0, \frac{-\lambda_3 M_R - \lambda_1 \eta B}{2 C_2} \right\}, u_{2,max} \right\}. \label{eq:opt_u2}
\end{align}

These characterizations provide profound insight into the mechanics of the optimal cocktail. The soft treatment $u_1^*$ is driven entirely by the phoretic mite load $M_P$ and its shadow cost $-\lambda_2$. Conversely, the harsh treatment $u_2^*$ contains a direct penalty term ($-\lambda_1 \eta B$). When the marginal value of the bee population $\lambda_1(t)$ is high, the optimal strategy will strictly suppress the use of $u_2^*$ to prevent toxicity-induced collapse, seamlessly transitioning the burden of control to the soft treatment $u_1^*$.

\subsection{Results}
\begin{figure}[htbp]
    \centering
    \includegraphics[width=0.7\linewidth]{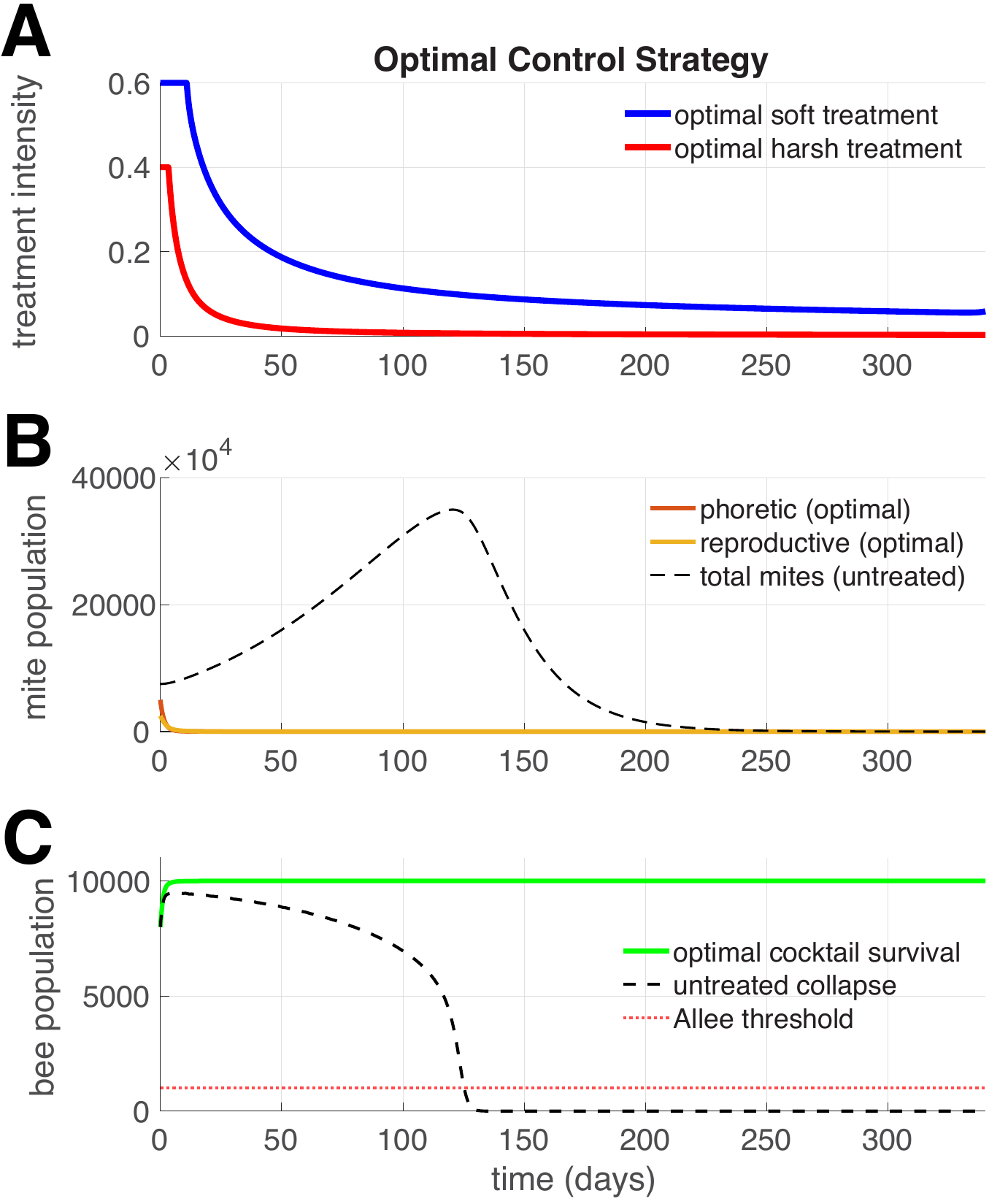}
    \caption{Optimal control profiles for $u_1(t)$ and $u_2(t)$ and the resulting mite and bee dynamics. (A) Optimal treatment protocols; (B) Resulting mite dynamics; (C) Resulting bee dynamics. Parameter values are given in Tables~\ref{tab:parameters} except we take $K = 10000$ and $A = 1000$. We take $\eta = 0.05$}.
    \label{fig:optimal_control}
\end{figure}

To determine the most effective dynamic treatment protocol, we numerically solved the optimality system using a forward-backward sweep method. The resulting optimal control profiles, alongside the corresponding population trajectories, are presented in Figure~\ref{fig:optimal_control}.

The numerical solution fundamentally validates the theoretical constraints identified in our preceding bifurcation analysis. The optimal strategy dictates an immediate, maximal application of both the soft ($u_1^*$) and harsh ($u_2^*$) treatments to rapidly break the infestation cycle (Figure~\ref{fig:optimal_control}A). However, the temporal trajectories of the two controls rapidly diverge to balance efficacy against physiological cost.

Because the harsh treatment introduces sublethal toxicity ($-\eta u_2 B$), prolonged application risks shifting the Allee threshold and driving the colony into the saddle-node catastrophe. The optimal controller naturally avoids this fold by utilizing $u_2^*$ strictly as an acute shock. It penetrates the capped brood to clear the reproductive reservoir ($M_R$), but the dosage is exponentially decayed to near-zero within the first 50 days to spare the healthy bees. 

In contrast, the soft treatment ($u_1^*$) incurs no toxic penalty. Following the initial crisis phase, the optimal strategy transitions to a continuous, low-intensity application of $u_1^*$. This chronic maintenance dose effectively holds the transcritical threshold ($B^*_{soft}$) safely above the natural carrying capacity $K$, permanently suppressing the phoretic mites ($M_P$) without penalizing hive growth.

The ecological efficacy of this ``shock and maintain'' cocktail is stark. In the untreated regime, the parasite population undergoes a massive explosion, ultimately driving the host bee population down through the unstable saddle point $E_A$ and triggering a complete collapse (dashed trajectories in Figure~\ref{fig:optimal_control}B and C). Under the optimal control protocol, the mites are asymptotically eradicated, allowing the recovering hive to rapidly stabilize at its optimal carrying capacity $K$.


\section{Emergence of Treatment-Resistant Mites}

The continuous application of single-target miticides has inadvertently selected for resistant strains of \textit{Varroa destructor}. To understand the impact of resistance on hive viability, we expand our model to include a parallel population of treatment-resistant mites, acknowledging that resistance primarily emerges as a mutation during the reproductive cycle.


Let $M_{P,s}$ and $M_{R,s}$ denote the susceptible phoretic and reproductive mites, respectively, and let $M_{P,r}$ and $M_{R,r}$ denote the resistant strain. We introduce a mutation probability $\mu \in (0, 1)$, which represents the fraction of offspring emerging from susceptible reproductive mites that acquire resistance. We assume the resistant strain is completely immune to the soft treatment ($u_1$) and exhibits a reduced susceptibility to the harsh treatment ($u_2$), defined by an efficacy parameter $\rho \in (0, 1)$.

The expanded five-dimensional system is:
\begin{align}
    \frac{dB}{dt} &= \alpha B(B-A)(K-B) - \beta B (M_{P,s} + M_{P,r}) - \eta u_2 B, \label{eq:res_B} \\
    \frac{dM_{P,s}}{dt} &= (1-\mu)\nu\sigma M_{R,s} - \gamma B M_{P,s} - (\delta + u_1) M_{P,s}, \label{eq:res_Mps} \\
    \frac{dM_{R,s}}{dt} &= -(\sigma + u_2) M_{R,s} + \gamma B M_{P,s}, \label{eq:res_Mrs} \\
    \frac{dM_{P,r}}{dt} &= \nu\sigma M_{R,r} + \mu\nu\sigma M_{R,s} - \gamma B M_{P,r} - \delta M_{P,r}, \label{eq:res_Mpr} \\
    \frac{dM_{R,r}}{dt} &= -(\sigma + \rho u_2) M_{R,r} + \gamma B M_{P,r}. \label{eq:res_Mrr}
\end{align}

\subsection{Equilibrium Shift and the Source-Sink Dynamic}
The introduction of the mutation rate $\mu$ fundamentally alters the equilibrium landscape. For the susceptible strain, setting the derivatives to zero yields a new critical bee population required to sustain the infestation:
\begin{equation}
    B^*_{s} = \frac{\delta + u_1}{\gamma \left( \frac{(1-\mu)\nu\sigma}{\sigma + u_2} - 1 \right)}.
\end{equation}
Because $(1-\mu) < 1$, the required host density $B^*_{s}$ is strictly larger than in the mutation-free model. The susceptible strain is weakened because a fraction of its reproductive effort is siphoned off to produce resistant mites.

However, the dynamics of the resistant strain are now governed by a source-sink relationship. Setting the derivative of the resistant phoretic mites to zero yields:
\begin{equation}
    0 = M_{P,r} \left[ \gamma B \left( \frac{\nu\sigma}{\sigma + \rho u_2} - 1 \right) - \delta \right] + \mu\nu\sigma M_{R,s}.
\end{equation}
The term $\mu\nu\sigma M_{R,s}$ acts as a continuous, strictly positive source term as long as the susceptible population is active. Consequently, the resistant strain no longer undergoes an independent transcritical bifurcation; it cannot be eradicated while the susceptible strain persists. 

To clear the hive entirely, the beekeeper must apply a static dosage of $u_2$ sufficient to force the denominators of both critical thresholds to be negative. Because $\rho < 1$, this requires an exceedingly high static dose of the harsh treatment. As demonstrated previously, this level of constant toxicity inevitably depresses the effective carrying capacity and triggers a saddle-node annihilation of the hive, necessitating a dynamic optimal control approach.

\section{Optimal Control with Resistant Mites}

\subsection{Objective Functional and Hamiltonian}
The objective remains to maximize bee survival while minimizing the total parasitic load and treatment costs across both strains. The objective functional is:
\begin{equation}
    J(u_1, u_2) = w_1 B(T) - \int_0^T \Big( w_2 (M_{P,s} + M_{P,r}) + w_3 (M_{R,s} + M_{R,r}) + C_1 u_1^2 + C_2 u_2^2 \Big) dt.
\end{equation}

We define the Hamiltonian $H$ by introducing five adjoint variables $\lambda_1, \dots, \lambda_5$:
\begin{align}
    H = &-\Big( w_2 (M_{P,s} + M_{P,r}) + w_3 (M_{R,s} + M_{R,r}) + C_1 u_1^2 + C_2 u_2^2 \Big) \nonumber \\
    &+ \lambda_1 \Big[ \alpha B(B-A)(K-B) - \beta B (M_{P,s} + M_{P,r}) - \eta u_2 B \Big] \nonumber \\
    &+ \lambda_2 \Big[ (1-\mu)\nu\sigma M_{R,s} - \gamma B M_{P,s} - (\delta + u_1) M_{P,s} \Big] \nonumber \\
    &+ \lambda_3 \Big[ -(\sigma + u_2) M_{R,s} + \gamma B M_{P,s} \Big] \nonumber \\
    &+ \lambda_4 \Big[ \nu\sigma M_{R,r} + \mu\nu\sigma M_{R,s} - \gamma B M_{P,r} - \delta M_{P,r} \Big] \nonumber \\
    &+ \lambda_5 \Big[ -(\sigma + \rho u_2) M_{R,r} + \gamma B M_{P,r} \Big]. \label{eq:hamiltonian_res}
\end{align}

By differentiating the Hamiltonian with respect to the state variables, we obtain the dynamic evolution of the adjoint system $\dot{\lambda}_i = -\frac{\partial H}{\partial x_i}$:
\begin{align}
    \frac{d\lambda_1}{dt} &= -\lambda_1 \Big[ \alpha(-3B^2 + 2(A+K)B - AK) - \beta(M_{P,s} + M_{P,r}) - \eta u_2 \Big] \nonumber \\
    &\quad + \gamma M_{P,s}(\lambda_2 - \lambda_3) + \gamma M_{P,r}(\lambda_4 - \lambda_5), \\
    \frac{d\lambda_2}{dt} &= w_2 + \lambda_1 \beta B + \lambda_2(\gamma B + \delta + u_1) - \lambda_3 \gamma B, \\
    \frac{d\lambda_3}{dt} &= w_3 - \lambda_2 (1-\mu)\nu\sigma - \lambda_4 \mu\nu\sigma + \lambda_3(\sigma + u_2), \\
    \frac{d\lambda_4}{dt} &= w_2 + \lambda_1 \beta B + \lambda_4(\gamma B + \delta) - \lambda_5 \gamma B, \\
    \frac{d\lambda_5}{dt} &= w_3 - \lambda_4 \nu\sigma + \lambda_5(\sigma + \rho u_2).
\end{align}
The transversality conditions at $t=T$ are $\lambda_1(T) = w_1$, and $\lambda_i(T) = 0$ for $i \in \{2,3,4,5\}$. Note that the dynamic shadow price of the susceptible reproductive mites ($\lambda_3$) is now explicitly penalized by the shadow price of the resistant phoretic mites ($\lambda_4$) due to the mutation linkage $\mu$.

Applying Pontryagin's Maximum Principle, we evaluate $\frac{\partial H}{\partial u_1} = 0$ and $\frac{\partial H}{\partial u_2} = 0$, yielding the optimal control characterizations bounded by $[0, u_{max}]$:
\begin{align}
    u_1^*(t) &= \min \left\{ \max \left\{ 0, \frac{-\lambda_2 M_{P,s}}{2 C_1} \right\}, u_{1,max} \right\}, \label{eq:opt_u1_res} \\
    u_2^*(t) &= \min \left\{ \max \left\{ 0, \frac{-\lambda_1 \eta B - \lambda_3 M_{R,s} - \lambda_5 \rho M_{R,r}}{2 C_2} \right\}, u_{2,max} \right\}. \label{eq:opt_u2_res}
\end{align}

\subsection{Results.}

\begin{figure}[htbp]
    \centering
    \includegraphics[width=0.7\linewidth]{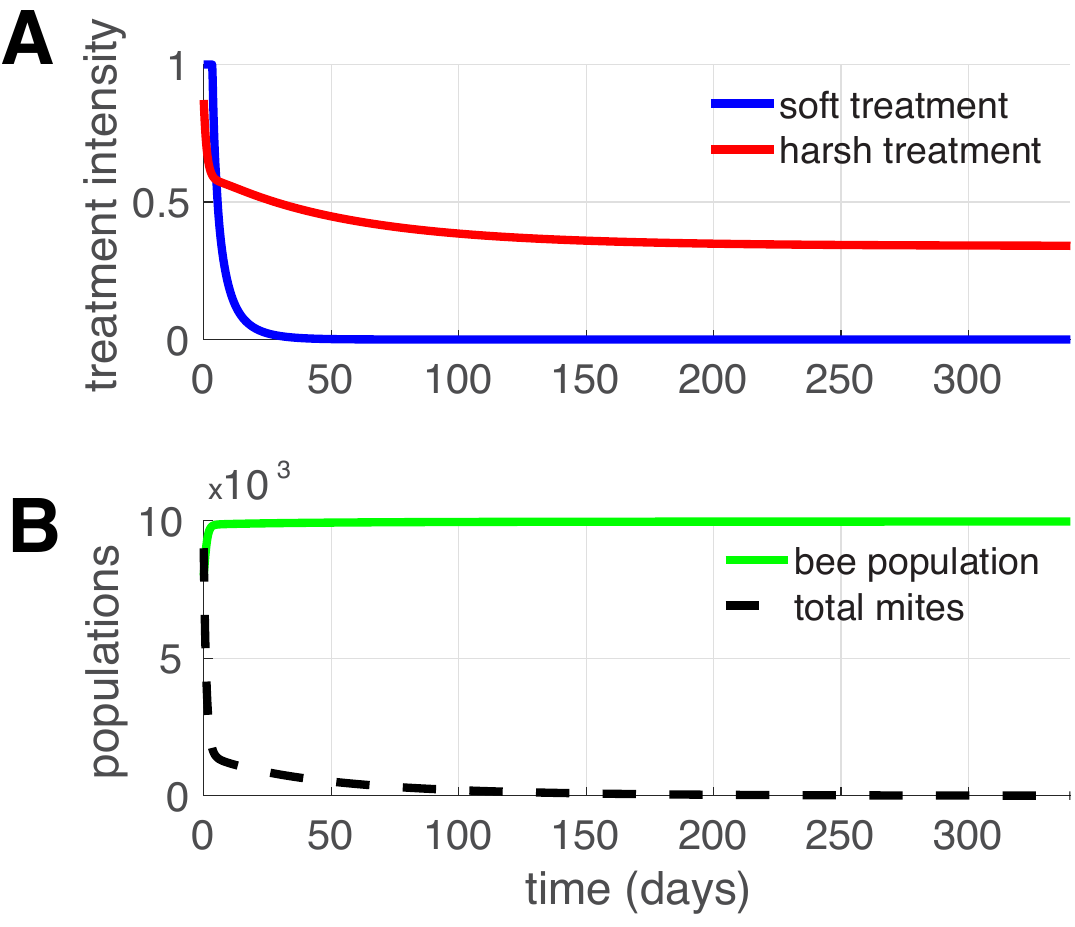}
    \caption{Optimal control profiles for $u_1(t)$ and $u_2(t)$ and the resulting mite and bee dynamics with resistant mites. (A) Optimal treatment protocols; (B) Resulting population dynamics.Parameter values are given in Table~\ref{tab:parameters} except we set $K = 10000$ and $A = 1000$. We set $\eta = 0.05$, $\mu = 0.02$, and $\rho = 0.4$.}
    \label{fig:optimal_control_res}
\end{figure}

Numerical solutions to the optimality system for the expanded resistance model reveal a drastic paradigm shift in the optimal treatment strategy (Figure~\ref{fig:optimal_control_res}). When a resistant subpopulation ($M_{P,r}, M_{R,r}$) is seeded into the initial conditions, the previously optimal ``shock and maintain'' cocktail fails.

Because the resistant strain is completely immune to the soft treatment, applying $u_1^*$ yields diminishing returns. Consequently, after a brief initial spike to clear the susceptible mites, the soft treatment is entirely abandoned ($u_1^* \to 0$). Prevention of the resistant mites from driving the colony to collapse relies exclusively on the harsh treatment. As seen in Figure~\ref{fig:optimal_control_res}A, $u_2^*(t)$ settles into a massive, chronic application rate ($u_2^* \approx 0.35$) for the remainder of the time interval. 

This numerical output perfectly mirrors the ``chemical treadmill'' observed in commercial apiculture: as resistance emerges, beekeepers are forced to apply increasingly toxic dosages of single-target miticides year-round simply to maintain baseline colony survival. While the hive population mathematically stabilizes at $K$ under this intense regimen (Figure~\ref{fig:optimal_control_res}B), the colony is locked into a state of severe chemical dependency.

Crucially, the reliance on high-dose harsh treatments places the hive in extreme ecological peril when seasonal dynamics are considered. As the hive transitions into winter or faces severe forage scarcity, the effective carrying capacity $K$ drops significantly. For instance, if environmental stressors reduce the carrying capacity to overwintering levels (e.g., $K = 1,500$), a profound ecological shift occurs: the hive drops below the transcritical establishment threshold ($B^* = 2,000$). Biologically, the shrinking winter cluster can no longer sustain the mite population, and the system naturally crosses the bifurcation boundary back to a stable, disease-free state. An adaptive optimal controller would recognize this structural shift and instantly withdraw all treatments.

However, commercial apiculture frequently relies on static, year-round ``chemical treadmills.'' If a beekeeper rigidly maintains the chronic summer dosage ($u_2 = 0.35$) into the autumn and winter months, the results are mathematically devastating. Recomputing the saddle-node catastrophe threshold (Eq.~\eqref{eq:u2crit}) for an overwintering hive ($K = 1,500, A = 1,000$) reveals that the boundary collapses to an infinitesimally small $u_{2,crit} \approx 0.0019$. Thus, blindly continuing the summer treatment protocol vastly exceeds the lethal limit for a winter hive. Any attempt to maintain this chemical treadmill year-round will push the marginalized colony far over the shifting Allee threshold, triggering irreversible hysteresis. This mathematically formalizes why hives heavily treated for resistant mites are highly susceptible to sudden overwintering losses: the treatment is not only highly toxic, but dynamically unnecessary.

\section{Conclusion}

The management of \textit{Varroa destructor} infestations in honey bee colonies represents a complex dynamical challenge, balancing the urgent need for parasite eradication against the physiological fragility of the hive. In this paper, we developed a mathematical model of hive-mite interactions incorporating a eusocial Allee effect and analyzed the efficacy of isolated and combined treatment strategies. 

Our equilibrium and bifurcation analyses reveal the fundamental inadequacy of static treatment paradigms. While soft (phoretic) treatments are safe for the bees, they fail to penetrate the capped brood reservoir, trapping the colony in a chronic endemic state. Conversely, harsh (reproductive) treatments effectively clear the reservoir but impose a severe toxicity penalty. We demonstrated analytically that applying a sufficiently high static dose of a harsh treatment permanently depresses the hive's carrying capacity and elevates the Allee threshold. This dynamic culminates in a catastrophic saddle-node bifurcation, where the attempt to cure the infestation mathematically guarantees colony collapse. 

To navigate this constrained safe operating space, we formulated an optimal control problem to derive a dynamic, time-dependent "cocktail" strategy. Using Pontryagin's Maximum Principle, we characterized an adaptive therapy protocol that relies on the soft treatment for continuous phoretic suppression, while reserving the toxic harsh treatment for precise, targeted pulses. The optimal controller inherently functions as an Allee collision-avoidance system, actively suppressing harsh treatments when the marginal value of the bee population dictates that the risk of toxicity-induced collapse outweighs the benefit of reservoir clearance. Furthermore, we demonstrated that the emergence of treatment-resistant mite strains drastically narrows the survival region of parameter and control space. When resistance emerges, the optimal control framework proves that temporal modulation is strictly necessary to prevent the hive from being forced onto an entirely lethal chemical treadmill. 

Beyond chemical interventions, our bifurcation analysis yields a striking ecological prediction regarding the natural mitigation of \textit{Varroa} and mite infestations. The mathematical boundary separating the disease-free state from the endemic regime is linked to the hive's carrying capacity. Analytically, the most natural path to keeping a colony mite-free is to maintain a smaller hive. Therefore, our model suggests a possible evolutionary consequence: if mite infestations continue to grow uncontrollably, evolutionary pressures will likely select for smaller honey bee colonies, favoring frequent swarming behaviors over the massive, static population sizes demanded by commercial apiculture.

While this theoretical framework provides rigorous insights into treatment timing and toxicity limits, we acknowledge several limitations. Our model relies on deterministic ordinary differential equations with constant parameters, which abstracts away the inherent stochasticity of weather, foraging availability, and discrete seasonal brood cycles. Additionally, we do not explicitly model the spatial dynamics of the hive or the transmission of secondary viral infections (such as Deformed Wing Virus), which often act as the ultimate proximal cause of colony death. Future iterations of this model could incorporate seasonal forcing to map how the saddle-node catastrophe boundary shifts dynamically throughout the year.

Ultimately, this work highlights the critical intersection of dynamical systems theory and pest management. It underscores a stark ecological reality: chemical interventions are not a consequence-free panacea. The persistent accumulation of sublethal miticides fundamentally alters the viability landscape of the hive, proving that aggressive treatment carries an intrinsic risk of mathematically guaranteeing the colony's demise. As the statistician George Box famously noted, ``all models are wrong, but some are useful.'' By translating complex dynamical thresholds into actionable, optimal control strategies, it is our hope that this model proves useful for beekeepers and agricultural policymakers seeking to break the chemical treadmill and secure the future of global pollination.


\bibliographystyle{siam}

\end{document}